\documentclass[12pt,letterpaper]{article}
\usepackage{graphicx}
\usepackage{amssymb,amscd}
\usepackage[latin1]{inputenc}
\usepackage{amsmath}
\usepackage{amsfonts}
\usepackage{amssymb}
\usepackage{graphicx,color}
\usepackage{young}
\usepackage[vcentermath]{youngtab}

\newcommand{\ie}{i.e.,~}

\newcommand{\ep}{\epsilon}

\newcommand{\be}{\begin{equation}}
\newcommand{\ee}{\end{equation}}

\begin{document}

\thispagestyle{empty}

\setcounter{page}{0}

\mbox{}

\begin{center} {\bf \Large  Duality-invariant bimetric formulation of linearized gravity}

\vspace{1.6cm}

Claudio Bunster$^{1,2}$,  Marc Henneaux$^{1,3}$ and Sergio H\"ortner$^3$

\footnotesize
\vspace{.6 cm}

${}^1${\em Centro de Estudios Cient\'{\i}ficos (CECs), Casilla 1469, Valdivia, Chile}

\vspace{.1cm}

${}^2${\em Universidad Andr\'es Bello, Av. Rep\'ublica 440, Santiago, Chile}

\vspace{.1cm}

${}^3${\em Universit\'e Libre de Bruxelles and International Solvay Institutes, ULB-Campus Plaine CP231, B-1050 Brussels, Belgium} \\

\vspace {15mm}

\end{center}

\centerline{\bf Abstract}
\vspace{.6cm}
A formulation of linearized gravity which is manifestly invariant under electric-magnetic duality rotations in the internal space of the metric and its dual, and which contains both metrics as basic variables (rather than the corresponding prepotentials), is derived.  In this bimetric formulation, the variables have a more immediate geometrical significance,  but the action is non-local in space, contrary to what occurs in the prepotential formulation.  More specifically, one finds that: (i) the kinetic term is non-local in space (but local in time); (ii) the Hamiltonian is local in space and in time; (iii)  the variables are subject to two Hamiltonian constraints, one for each metric.

Based in part on the talk ``Gravitational electric-magnetic duality" given by one of us (MH) at the 8-th Workshop ``Quantum Field Theory and Hamiltonian Systems" (QFTHS), 19-22 September 2012, Craiova, Romania.  To appear in the Proceedings of the Conference (special issue of the Romanian Journal of Physics).

\vspace{.8cm}
\noindent

\newpage

\section{Introduction}

\setcounter{equation}{0}

Understanding gravitational duality is one of the important challenges for exhibiting the conjectured infinite-dimensional Kac-Moody algebras (or generalizations thereof) of hidden symmetries of supergravities and M-theory \cite{Julia:1982gx,West:2001as,Damour:2002cu,Henneaux:2010ys}.   Independently of the problem of uncovering these conjectured hidden symmetries,  gravitational duality is important in itself as it illuminates the structure of Einstein gravity.

In \cite{Henneaux:2004jw}, two of the present authors presented a formulation of linearized gravity in four space-time dimensions that was manifestly invariant under ``duality rotations" in the internal space spanned by the graviton and its dual.  This was followed by further developments covering higher spins \cite{Deser:2004xt},  the inclusion of a cosmological constant  \cite{Julia:2005ze} and supersymmetry \cite{Bunster:2012jp}. 

One crucial aspect of the manifestly duality-invariant formulation of \cite{Henneaux:2004jw} was the introduction of prepotentials. Technically, these prepotentials arise through the solutions of the constraints present in the Hamiltonian formalism. The prepotential for the metric (spin-2 Pauli-Fierz field in the linearized theory) appears through the solution of the Hamiltonian constraint, while the prepotential for its conjugate momentum appears through the solution of the momentum constraint.  Explicitly, if $h_{ij}$ are the spatial components of the metric deviations from flat space and $\pi^{ij}$ the corresponding conjugate momenta, one has
\begin{equation}
h_{ij} = \ep_{irs} \partial^r \Phi^{s}_{\; \; j} + \ep_{jrs} \partial^r \Phi^{s}_{\; \; i} + \partial_i u_j + \partial_j u_i \label{hPhi0}
\end{equation}
and
\begin{equation}
\pi^{ij} = \ep^{ipq} \ep^{jrs} \partial_p \partial_r P_{qs} \label{piP}
\end{equation}
where $\Phi_{rs} = \Phi_{sr}$ and $P_{rs} = P_{sr}$ are the two prepotentials (the vector $u_i$ can also be thought of as a prepotential but it drops out from the theory so that we shall not put emphasis on it).   

The second metric $f_{ij}$ dual to $h_{ij}$ is defined in terms of the second prepotential $P_{ij}$ exactly as $h_{ij}$ is defined in terms of $\Phi_{ij}$,
\begin{equation}
f_{ij} = \ep_{irs} \partial^r P^{s}_{\; \; j} + \ep_{jrs} \partial^r P^{s}_{\; \; i} + \partial_i v_j + \partial_j v_i ,\label{fP0}
\end{equation}
the vector $v_i$ being another prepotential, which, just as $u_i$, drops from the theory.   The expressions (\ref{hPhi0}) and (\ref{fP0}) satisfy identically the Hamiltonian constraints,
\begin{equation}
R[h]  = 0 \label{H1constraint}
\end{equation}
and
\begin{equation}
R[f]  = 0 \label{H2constraint}
\end{equation}
where $R[h]$ and $R[f]$ are respectively the three-dimensional spatial curvatures of $h_{ij}$ and $f_{ij}$.  Explicitly,
$$ R[h] = \partial^m \partial^n h_{mn} - \triangle h, \; \; \; \;  R[f] = \partial^m \partial^n f_{mn} - \triangle f, $$
where $h$ and $f$ are the traces of $h_{mn}$ and $f_{mn}$, \ie $h = h_i^{\  i}$, $f = f_{i}^{\ i}$ and where $\triangle$ is the Laplacian.

When reformulated in terms of the prepotentials, duality symmetry  simply amounts to $\textrm{SO}(2)$ rotations in the internal plane of the prepotentials and consequently, also to $\textrm{SO}(2)$ rotations in the internal plane of the metrics.   The temporal components of the metrics arise through the integration of the equations of motion, as arbitrary ``integration functions".  The equations of motion can furthermore be interpreted as twisted self-duality conditions on the curvature tensors of the graviton and its dual \cite{Bunster:2012km}.   [For some background information on twisted self-duality, see \cite{Cremmer:1998px,Hull:2001iu,Bunster:2011qp}.]

The prepotentials are necessary for  locality of the action principle but  do not have (yet?) an immediate geometrical interpretation.  The metrics appear in this formulation as secondary.  
The purpose of this note is to provide a manifestly duality-invariant formulation of the theory in which the metrics $h_{ij}$ and $f_{ij}$ are the basic variables.  As we explicitly show in the next section, this is possible, but the price paid is non-locality of the action principle.

\section{Bimetric formulation}
\setcounter{equation}{0}

The key to the bimetric formulation of the variational principle relies on the fact that one can invert the relations (\ref{hPhi0}) and (\ref{fP0}) to express, up to gauge transformation terms for $\Phi_{ij}$ and $P_{ij}$ that drop from the action, the prepotentials in terms of the metrics when these latters satisfy the Hamiltonian constraints (\ref{H1constraint}) and (\ref{H2constraint}).  Remarkably enough, the expression takes almost the same form and read
\begin{equation}
\Phi_{ij} =- \frac{1}{4} \left[ \ep_{irs} \triangle^{-1} \left(\partial^r  h^{s}_{\; \; j} \right)+ \ep_{jrs}  \triangle^{-1} \left( \partial^r h^{s}_{\; \; i} \right) \right] \label{Phih1}
\end{equation}
and
\begin{equation}
P_{ij} = - \frac{1}{4} \left[ \ep_{irs} \triangle^{-1} \left(\partial^r  f^{s}_{\; \; j} \right)+ \ep_{jrs}  \triangle^{-1} \left( \partial^r f^{s}_{\; \; i} \right) \right]. \label{Pf1}
\end{equation}
One may easily verify that if one replaces (\ref{Phih1}) and (\ref{Pf1}) in (\ref{hPhi0}) and (\ref{fP0}) and uses the constraints (\ref{H1constraint}) and (\ref{H2constraint}), one recovers indeed $h_{ij}$ and $f_{ij}$, with some definite $u_i$ and $v_i$ that are not needed here and will not be written explicitly.

The expressions for the prepotentials are of course not unique since these are determined by the metrics up to prepotential gauge transformations, which have been analyzed in  \cite{Henneaux:2004jw} and which do not matter for our purposes since the theory is gauge invariant. The expressions (\ref{Phih1}) and (\ref{Pf1}) correspond to a specific choice of gauge. 

We can now substitute (\ref{Phih1}) and (\ref{Pf1}) in the manifestly duality action of \cite{Henneaux:2004jw}.  The ``$p$-$\dot{q}$"- term $K[Z_a^{\; mn}]$, 
\begin{equation}
K[Z_a^{\; mn}] = \int dt  \int d^3x \, \ep^{ab} \ep^{mrs} \left(\partial^p \partial^q \partial_r Z_{aps} - \triangle \partial_r Z_{a\;   s}^{\;q}\right) \dot{Z}_{bqm}  \label{Kinetic0}
\end{equation}
($a, b = 1,2$) where $(Z^a_{ij}) \equiv (P_{ij},\Phi_{ij})$ becomes
\begin{equation}
K[h^a_{\; mn}] = \frac{1}{4}\int dt  \int d^3x \, \ep_{ab} \ep^{mij} \left(\partial_i h^{an}_{\ \ \  j} - \partial^n \partial^r \partial_i \triangle^{-1}(h^{a}_{\ \ r  j})\right) \dot{h}^b_{\; mn}   \label{Kinetic1}
\end{equation}
with $(h^a_{ij}) \equiv (f_{ij},h_{ij})$.  It is evidently non-local in space (but local in time).
By contrast, the Hamiltonian is local and reads
\begin{eqnarray}
H[h^a_{\; mn}] &=& \int d^3x \delta_{ab} \left[ \frac{1}{4} \partial^r h^{a mn} \partial_r h^{b}_{\  mn} - \frac{1}{2} \partial_m h^{a m}_{\ \ \ \ n} \partial_r h^{brn}  \right] \nonumber \\ && + \int d^3x \delta_{ab} \left[\frac{1}{2} \partial^m h^a \partial^n h^{b}_{\ mn}  - \frac{1}{4} \partial^m h^{a} \partial_m h^{b}   \right] .
\end{eqnarray}  
Both the kinetic term and the Hamiltonian are invariant under linearized diffeomorphisms,
\begin{equation}
\delta h^a_{mn} = \partial_m \xi^a_n + \partial_n \xi^a_m
\end{equation}
up to surface terms.

Since the metrics are subject to the constraints (\ref{H1constraint}) and (\ref{H2constraint}), one must add these constraints to the action with Lagrange multipliers $n^a$ ($a=1,2$) when trading the prepotentials for the metrics.  The Lagrange multipliers turn out to be the linearized lapses of each metric.  The bimetric action is therefore
\begin{equation}
S[h^a_{\; mn}, n^a] = K[h^a_{\; mn}] - \int dx^0 H[h^a_{\; mn}] - \int dx^0 d^3x \delta_{ab} n^a R^b \label{action2}
\end{equation}
and is clearly invariant under $\textrm{SO}(2)$-duality rotations in the internal plane of the metrics,
\begin{equation}
f'_{ij} = \cos \alpha f_{ij} - \sin \alpha h_{ij}
\end{equation}
\begin{equation}
h'_{ij} = \sin \alpha f_{ij} + \cos \alpha h_{ij},
\end{equation}
accompanied with a similar rotation for the lapses $n^1$ and $n^2$.  This is because both $\epsilon_{ab}$ and $\delta_{ab}$ are invariant tensors.  In (\ref{action2}), $R^a$ stands for $R[h^a]$.

Note finally that the prepotentials are each worth two independent physical functions since they are unconstrained but possess four independent gauge symmetries (3 spatial diffeomorphisms and 1 Weyl rescaling). This gives $6$ (number of prepotential components) $- 4$ (number of prepotential gauge symmetries) $=2$ physical functions.  The metrics each contain the same number of independent physical components, as they should.  They are subject to one constraint  (Hamiltonian constraint) but possess only three independent gauge symmetries (spatial diffeomorphisms).  This gives again $6$ (number of metric components) $-1$ (number of constraints) $-3$ (number of metric gauge symmetries) $=2$ physical functions.  The conjugate momentum also contains two independent physical functions, but the counting is this time $6$ (number of momentum components) $-3$ (number of momentum constraints) $-1$ (number of momentum gauge symmetries) $=2$ physical functions.

\section{Conclusions and comments}
The manifestly duality-invariant metric action (\ref{action2}) is our main result.  Although it involves two metrics, we stress that this action is strictly equivalent to the linearized Einstein-Hilbert, or Pauli-Fierz, action.  In particular, it contains no additional massive or massless spin-2 degree of freedom. 

The action clearly exhibits that the two metrics are not only duality conjugate, but also canonically conjugate (in a generalized sense taking into account the non trivial, c-number, operator present in the kinetic term that takes the schematic form $\int dx^0 a_{AB} z^A \dot{z}^B$ where $A,B$ run over all discrete ($a$, $(m,n)$) and continuous ($\vec{x}$) indices, and where $a_{AB}$ is a c-number infinite square matrix which is not the standard symplectic matrix  $\begin{pmatrix} 0 & -I \\I & 0 \end{pmatrix}$ - it is actually degenerate -, so that the Poisson brackets are not the canonical ones).  The detailed Hamiltonian structure will be worked out and studied elsewhere \cite{BHH}.

The bimetric formulation is nonlocal in space (but local in time).  This might not really be a drawback since it is expected that the manifestly duality invariant formulation of the interacting theory (if it exists) will be non-local anyway.  The formulation underplays the role of the prepotentials, which are absent in (\ref{action2}).  The prepotentials are nevertheless useful technical devices which enable one to go from the conjugate momenta $\pi^{mn}$ to the second metric $f_{mn}$, and to control also the non-locality of the action.

Although the theory is  Poincar\'e invariant, the formulation lacks manifest space-time covariance.  This seems to be an unavoidable feature whenever one deals with manifest duality invariance \cite{Deser:1976iy,Henneaux:1988gg,Schwarz:1993vs}.   This might indicate that space-time covariance is somewhat secondary.  This point of view would seem  in any event to be inevitable if spacetime itself is a derived, emergent concept. There exist in fact models in which Poincar\'e invariance  can actually be derived  from duality invariance  \cite{Bunster:2012hm}.  

In higher dimensions, the dual to the Pauli-Fierz spin-2 field is a mixed Young-symmetry tensor field described by the Curtright action \cite{Curtright:1980yk}.  The two-field action analogous to (\ref{action2}) involves simultaneously in that case the standard graviton $h_{mn}$ and the Curtright field $T_{m_1 \cdots m_{D-3} n}$. The details will be given elsewhere \cite{BHH}.

\section*{Acknowledgments} 
M.H. is grateful to the organizers of the 8-th Workshop ``Quantum Field Theory and Hamiltonian Systems" (QFTHS, Craiova, Romania) for their kind invitation. C.B. and M.H.  thank  the Alexander von Humboldt Foundation for Humboldt Research Awards.  The work of M.H.and S.H. is partially supported by the ERC through the ``SyDuGraM" Advanced Grant, by IISN - Belgium (conventions 4.4511.06 and 4.4514.08) and by the ``Communaut\'e Fran\c{c}aise de Belgique" through the ARC program.  The Centro de Estudios Cient\'{\i}ficos (CECS) is funded by the Chilean Government through the Centers of Excellence Base Financing Program of Conicyt.

\end{document}